\documentclass[useAMS,usenatbib,psfig]{mn2e}
\usepackage{graphicx}
\begin{document}
\title[Nova Mon 2012]{Photometric evolution, orbital modulation and progenitor of Nova Mon 2012}

\author[U. Munari et al.]{U. Munari$^{1}$, S. Dallaporta$^{2}$, F. Castellani$^{2}$, P. Valisa$^{2}$, A. Frigo$^{2}$, L. Chomiuk$^{3}$,
\newauthor V.A.R.M. Ribeiro$^{4}$\\
$^{1}$INAF Astronomical Observatory of Padova, 36012 Asiago (VI), Italy\\
$^{2}$ANS Collaboration, c/o Osservatorio Astronomico, via dell'Osservatorio 8, 36012 Asiago (VI), Italy\\
$^{3}$MSU, Chicago\\
$^{4}$Astrophysics, Cosmology and Gravity Centre, Astronomy Dept., 
             Univ. of Cape Town, Private Bag X3, Rondebosch 7701, South Africa}

\date{Accepted .... Received ....; in original form ....}

\pagerange{\pageref{firstpage}--\pageref{lastpage}} \pubyear{2010}

\maketitle

\label{firstpage}

\begin{abstract}
We present and discuss accurate and densely mapped $B$$V$$R_{\rm C}$$I_{\rm
C}$ lightcurves of the neon Nova Mon 2012, supplemented by the evolution in
Stromgren $b$ and $y$ bands and in the integrated flux of relevant emission
lines.  Our monitoring started with the optical discovery of the nova (50
days past the first detection in gamma-rays by Fermi-LAT) and extend to day
+270, well past the end of the super-soft phase in X-rays.  The nova was
discovered during the nebular decline, well past $t_3$ and the transition to
optically thin ejecta.  It displayed very smoothly evolving lightcurves.  A
bifurcation between $y$ and $V$ light-curves took place at the start of the
SSS phase, and a knee developed toward the end of the SSS phase.  The
apparent magnitude of the nova at the unobserved optical maximum is
constrained to +2.8$\leq$$V$$\leq$4.2.  The appearance, grow in amplitude
and then demise of a 0.29585 ($\pm$0.00002) days orbital modulation of the
optical brightness was followed along the nova evolution.  The observed
modulation, identical in phase and period with the analog seen in the X-ray
and satellite ultraviolet, has a near-sinusoidal shape and a weak secondary
minimum at phase 0.5.  We favor an interpretation in terms of
super-imposed ellipsoidal distortion of the Roche lobe filling companion and
irradiation of its side facing the WD.  Similar lightcurves are typical of
symbiotic stars where a Roche lobe filling giant is irradiated by a very hot
WD.  Given the high orbital inclination, mutual occultation between the
donor star and the accretion disk could contribute to the observed
modulation. The optical+infrared spectral energy distribution of Nova Mon
2012 during the quiescence preceeding the outburst is nicely fitted by a
early K-type main-sequence star ($\sim$K3V) at 1.5 kpc distance,
reddened by $E_{B-V}$=0.38, with a WD companion and an accretion disk
contributing to the observed blue excess and moderate H$\alpha$ emission.  A
typical early K-type main-sequence star with a mass of $\sim$0.75 M$_\odot$
and a radius of $\sim$0.8 R$_\odot$, would fill its Roche lobe for a
P=0.29585 day orbital period and a more massive WD companion (as implied by
the large Ne overabundance of the ejecta).
\end{abstract}
\begin{keywords}
Stars: novae 
\end{keywords}

\section{Introduction}

Nova Mon 2012 was discovered by S. Fujikawa on 2012 Aug 9.81 UT when it was
just emerging from conjunction with the Sun, only briefly observable low on
the eastern horizon at dawn.  First spectroscopic quick looks reported in  
CBET 3202 suggested the nova had aleardy significantly evolved past maximum
brightness.  Cheung et al.  (2012a) noted the position of the nova was
consistent with that of Fermi J0639+0548, being within the preliminary 68\%
confidence LAT error circle radius of 0.12 deg (statistical uncertainty
only).  Fermi J0639+0548 was a new gamma-ray transient in the Galactic 
plane, detected by Fermi Large Area Telescope from 2012 June 22 to 24, when
its proximity with the Sun prevented observations on that area of the sky  
from the ground or with X-ray satellites.  The positional proximity and the
temporal compatibility, prompted Cheung et al.  (2012) to suggest that Nova
Mon 2012 was observed as Fermi J0639+0548 at the time of initial outburst. 
A similar association between a classical nova and a gamma-ray source was  
proposed on firmer basis for Nova Sco 2012 (Cheung et al.  2012b), and     
firmly established for the 2010 outburst of V407 Cyg (Abdo et al.  2010).  
There are no other known classical or recurrent novae associated to
transient gamma-ray sources.

The tentive link of Nova Mon 2012 with Fermi J0639+0548 boosted the interest
in this object.  The first detailed optical observations were presented by
Munari et al.  (2012a) that observed the nova between Aug 16.14 and Aug
20.13 and presented accurate $B$$V$$R_{\rm C}$$I_{\rm C}$ photometry and
absolutely fluxed high resolution Echelle spectroscopy, indicating the nova
suffers from modest interstellar $E_{B-V}$+0.30 reddening, evolving through
its early nebular phase, with double peaked emission line profiles
characterized by FWHM ranging from 2010 to 3035 km/s.  The radial velocity
of interstellar absorption lines constrained between 1 and 2 kpc the
distance to the nova.  At later times, Munari (2013) found Nova Mon 2013 to
be a Neon nova from the marked over-abundance of Ne in the spectra,
dominated by [NeIII], [NeIV] and [NeV] emission lines, which indicates that
the white dwarf is massive and of the ONe type.  Nelson et al.  (2012a)
observed the nova with Swift on Aug 19 and found its spectrum to be hard and
highly absorbed, with the majority of the counts having energies above 2
keV.  Other X-ray observations taken on Sept 12 were reported by Ness et al. 
(2012) that indicated a thinning of the self-absorption by the ejecta and a
variability in the flux level.  Chomiuk et al.  (2012) reported about VLA
radio detection of the nova on June 30, well before optical rediscovery, and
its subsequent rise in radio brightness.  Combined with the eruption date of
the Fermi source taken as the time of ejection, the radio data constrain to
$\sim$1.4 kpc the distance to the nova.  Further radio data were presented
by Fuhrmann et al.  (2012).  Greimel et al.  (2012) noted how the progenitor
of Nova Mon 2012 was detected by IPHAS in 2004, well before the outburst, as
a star of $r'$=17.9, $r'$$-$$i'$=+0.69 with mild H$\alpha$ line emission and
short term variability.  The combined optical IPHAS and infrared UKIDSS
colors appear consistent with those of an early K-type star and
$E_{B-V}$+0.30, requiring instead a much larger reddening if the fit was
attempted with an optically thick accretion disk dominating at all
wavelengths.

O'Brien et al. (2012), obtained on Sept 18 interferometric images with the
European VLBI Network that resolved Nova Mon 2012 into two distinct sources
separated by $\sim$35 mas.  If these components were associated with
material ejected in the explosion supposed to have occoured at the time of
the eruption of the Fermi source, the separation indicates an angular
expansion of 0.4 mas/day. Super-soft X-ray emission was eventually detected
by Nelson et al. (2012b) for the first time on Swift observations for
November 11, in addition to the still present hard component, and found to
be highly variable on a short time scale by Li and Kong (2012). Page et al.
(2013a) reported about a rapid and deep decline of the SSS component during 
February 2013, and Page et al. (2013b) about the complete disappearance of
the SSS component by early March 2013.

Infrared photometric and spectroscopic observations were reported by
Banerjee et al.  (2012), Varricatt et al.  (2012), and Banerjee et al. 
(2013).  They traced high ionization emission lines from November 2012 to
January 2013 and confirmed the FWHM of emission lines found earlier in the
optical.  The rate of brightness decline in $J$$H$$K$ bands was similar to
that of optical bands.  Banerjee et al.  (2013), by comparing the 2MASS and
UKIDSS data for the progenitor of Nova Mon 2012, found $J$$H$$K$ variability
in quiescence ($\sim$0.5 mag amplitude), and suggested that excess in WISE
longer wavelength colors indicates the presence of an extended dusty
envelope around the progenitor.  To account for the initial production of
gamma-rays, Banerjee et al. proposed that the donor star in Nova Mon 2012
is a cool giant.

In this paper we discuss the photometric evolution of Nova Mon 2012 in broad
$B$$V$$R_{\rm C}$$I_{\rm C}$ bands, narrow Stromgren $b$ and $y$ bands, and
in the integrated flux of optical emission lines from the discovery to after
the end of the SSS phase. The orbital modulation and its variation with
wavelength and time are discussed, together with the irradiation of the
donor star during the SSS phase and subsequent evolutionary stages. The
nature of the progenitor and the distance to the system are derived by
combining pre-outburst optical and infrared photometry with the orbital
period and evidence of Roche lobe filling conditions.

  \begin{figure*}
    \centering   
    \includegraphics[width=17.5cm]{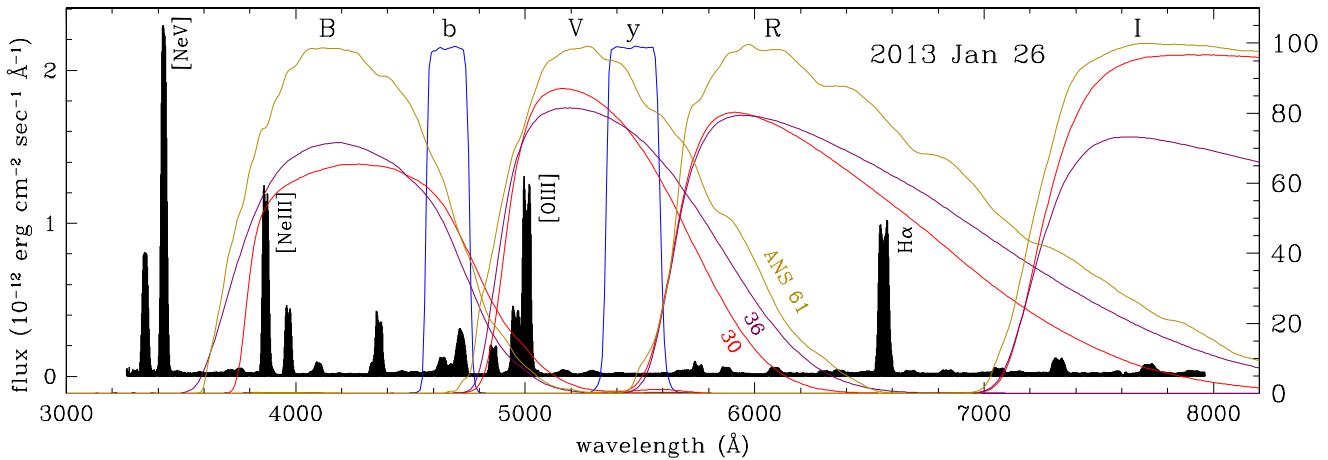}
     \caption{Transmissions of the actual filters sets of the three ANS
  Collaboration telescopes used to monitor the photometric evolution of Nova
  Mon 2012.  They are overplotted to a spectrum of the nova to highlight how
  different emission lines contribute to the flux observed in the various
  photometric bands.}
     \label{fig1}
  \end{figure*}

\section{Observations}

$B$$V$$R_{\rm C}$$I_{\rm C}$ photometry of Nova Mon 2012 was obtained with
three of the telescopes operated by ANS Collaboration (N.  30, 36 and 61). 
Technical details of this network of telescopes running since 2005, their
operational procedures and sample results are presented by Munari et al. 
(2012b).  Detailed analysis of the photometric performances and measurements
of the actual transmission profiles for all the photometric filter sets in
use is presented by Munari and Moretti (2012).  All measurements on Nova Mon
2012 were carried out with aperture photometry, the long focal length of the
telescopes and the absence of nearby contaminating stars not requiring the
use of PSF-fitting.  All photometric measurements were carefully tied to a
local $B$$V$$R_{\rm C}$$I_{\rm C}$ photometric sequence extracted from the
APASS survey (Henden et al.  2013) and ported to the Cousin and Stromgren
systems following transformation equations of Munari (2012).  The
photometric sequence was selected to densely cover a color range larger than
displayed by the nova during its evolution.  The sequence was intensively
tested during the whole observing campaing for linearity of color equations
and for absence of intrinsic variability of any of its constituent stars. 
The use of the same, high quality photometric comparison sequence for all
the involved telescopes and for all observations of the nova ensued the
highest internal homogeneity of the collected data.  The median value of the
total error budget (defined as the quadratic sum of the Poissonian error on
the nova and the formal error of the transformation from the local to the
standard system as defined by the local photometric sequence) is 0.007 mag
for $B$, 0.006 in $V$, 0.007 in $R_{\rm C}$, 0.008 in $I_{\rm C}$, and 0.008
mag for $B-V$, 0.008 in $V-R_{\rm C}$, and 0.009 in $V-I_{\rm C}$.  Colors
and magnitudes are obtained separately during the reduction process, and are
not derived one from the other.

  \begin{table}
    \centering      
     \caption{Journal of our spectroscopic observations of Nova Mon 2012.}
    \includegraphics[width=7.8cm]{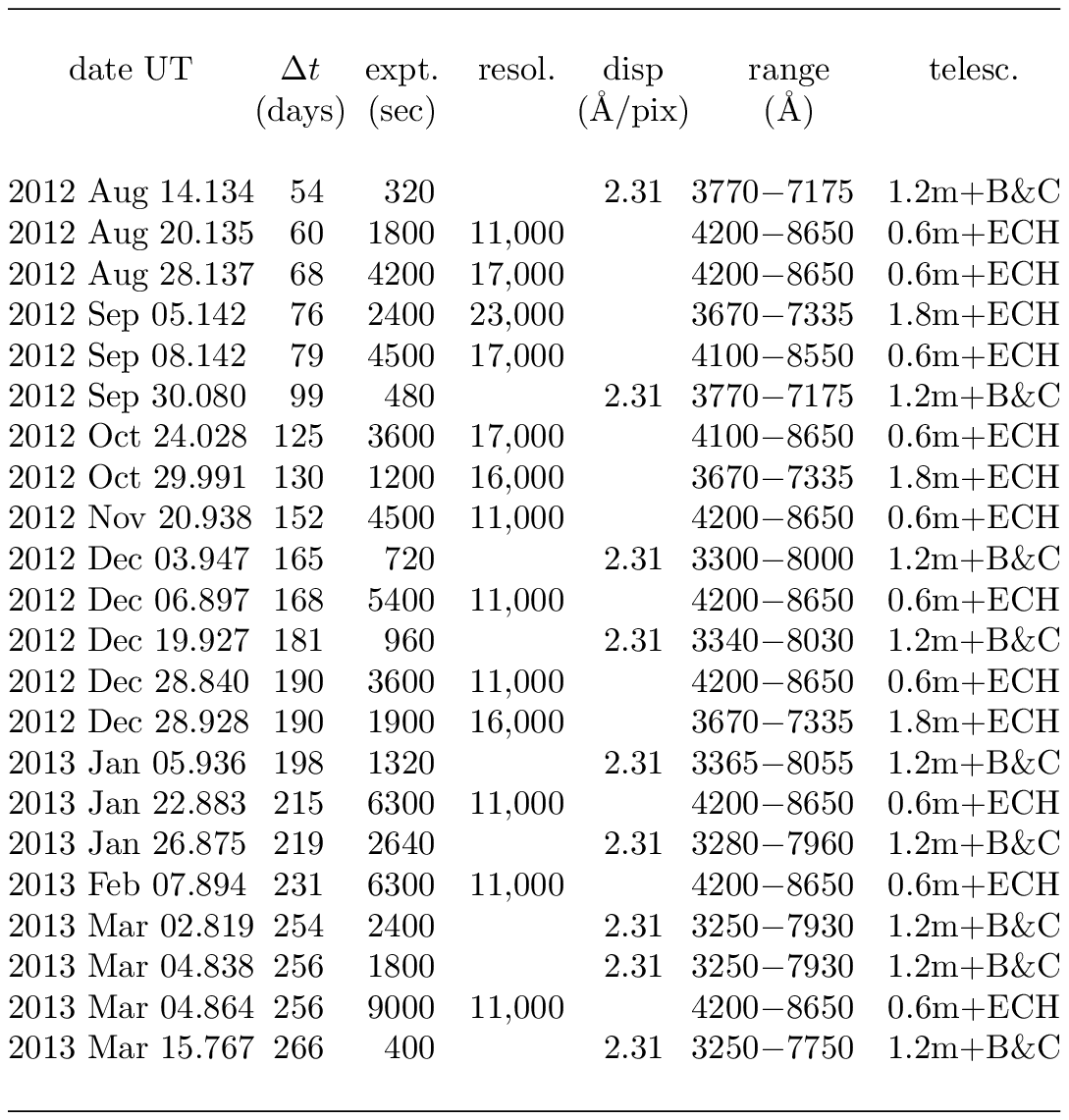}
     \label{tab1}
  \end{table}    

Spectroscopic observations of Nova Mon 2012 have been obtained with three
different telescopes.  High resolution Echelle spectra were provided by
Asiago 1.82m telescope operated by the National Institute of Astrophysics
(INAF), and by ANS Collaboration 0.6m telescope operated by Schiaparelli
Observatory in Varese.  Low resolution spectra were obtained with the B\&C
spectrograph mounted on the Asiago 1.22m telescope of the University of
Padova.  All spectra were obtained in long slit mode, with the slit rotated
to the parallactic angle, and about 2 arcsec in width.  All spectra were
calibrated into absolute fluxes, by observing immediately before and after
the nearby standard stars HR 1578 and HR 2315.  The high accuracy of the
fluxes were checked against the nearly simultaneous, daily optical
photometry.  Over the 3900-6800 \AA\ range, the error of the relative fluxes
in the finally reduced spectra does not exceed a few percent, and the error
of the global zero point never exceeds 7\%.  A log of the spectroscopic
observations is presented in Table~1.  In this paper the spectra will be
used for two main purposes: (a) extract convolved $y$-band Stromgren
photometry that trace the flux of the emission in the continuum between the
emission lines, (b) measure the integrated flux of selected emission lines. 
The information about the profiles of emission lines, their evolution with
time and how they constrain the geometry of the ejecta has been used by
Ribeiro et al.  (2013) in their 3D modeling of Nova Mon 2012 ejecta, and other 
information extracted from these spectra will be used by Chomiuk et al.
(2013) to support the interpretation and modeling of the radio evolution.  
One of the low resolution spectra we obtained is used as a reference 
background to Figure~1.

\section{Merging individual lightcurves}

Local realizations of a standard photometric system include different
atmospheric and filter transmission, detector sensitivity, efficiency of the
optics, etc.  On normal stars (like those making the photometric comparison
sequence) all these effects are collectively corrected for by the usual
transformation equations, provided that the local realization is
sufficiently close to the standard system, which is ensured by the
photometric filters having transmission profiles and the detector a
wavelength dependent sensityvity close to those used in defining the
standard system.  The spectra of novae, expecially during advanced decline,
when they are strongly dominated by broad and intense emission lines, are
completely different from the black-body like spectra of normal field stars,
and the transformation equations can only {\it partially} compensate for the
differences between the local and the standard systems.  The situation is
illustrated in Figure~1, where the measured transmission profiles for the
actual $B$$V$$R_{\rm C}$$I_{\rm C}$ filter sets used with the adopted ANS
telescopes are over-plotted to a spectrum of Nova Mon 2012 obtained during
the advanced decline (for details on how the transmission of filters of all
ANS telescope are measured and monitored over time see Munari and Moretti
2012).  These filters come from different manufacturers and are both of the
multi-layer dielectric type (Astrodon brand for ANS telescope N.61) and the
classical sandwich of colored glasses (Custom for telescope N.36, and
Schuler for N.30), built according to the standard recipe of Bessell (1990). 
It's obvious from Figure~1 how different filter sets transmit different
relative amounts of continuum and emission line flux.

The light-curves of the same nova obtained with different telescopes will
therefore show offsets among them, and these offsets will be time-dependent
following the spectral evolution of the nova, characterized by smooth and
monotonic trends in (1) the contrast between emission lines and underlying
continuum, and (2) the relative intensity of the emission lines. The closest
representation of the {\it true} light-curve appears to be a proper
combination of different individual light-curves, effectively averaging over
different local realizations of the same standard system.  To combine them
into a merged light-curve, we adopt the LMM method introduced by Munari et
al. (2013) when dealing with the light-curves of the recent supernovae SN
2011fe in M101, SN 2012aw in M95, and SN 2012cg in NGC 4424. In short, we
assume that all individual light-curves from different telescopes are exactly
the same one, affected only by an offset in the ordinate zero-point and a
linear stretch of the ordinates around a pivot point:
\begin{eqnarray}
\noindent
{\rm mag_{i,j}(\lambda)} &=& {\rm mag^\ast_{i,j}(\lambda)} + \theta_j(\lambda)
+ \\ \nonumber
&+& \phi_j(\lambda) \times [t_{i,j}(\lambda) - t_j^\circ(\lambda)]
\end{eqnarray}
where mag$^\ast_{i,j}(\lambda)$ is the $i$-th magnitude value in the
$\lambda$ photometric band from the $j$-th telescope, mag$_{i,j}(\lambda)$
is the corresponding value in the merged light-curve, $\theta_j(\lambda)$ is
the amount of zero-point shift and $\phi_j(\lambda)$ is coefficient of the 
linear stretching around the $t^\circ_j (\lambda)$ pivot point. The stretch
is therefore a linear function of time difference with the reference $t^\circ_j
(\lambda)$ epoch.

The optimal values for $\theta_j(\lambda)$, $\phi_j(\lambda)$ and $t^\circ_j
(\lambda)$ are derived by a $\chi^2$ minimization of the difference between 
the observed points and those defining the merged light-curve:
\begin{equation}
     \chi^2 = \sum_{i} \sum_{j} \frac{[{\rm mag^\ast_{i,j}(\lambda)}
              - {\rm mag_{i,j}(\lambda)}]^2}{[\epsilon_{i,j}(\lambda)]^2}
\end{equation}
where $j$ sums over the different telescopes, $i$ over the individual
measurements of each telescope, and $\epsilon_{i,j}(\lambda)$ is the total
error budget of each measurement.  Notice that the LMM method to construct a
merged light-curve from observations obtained with different telescopes does
not stretch the resulting light-curve in flux, nor time, and does not
individually alter any single observational point.

Our merged $B$$V$$R_{\rm C}$$I_{\rm C}$ photometry of Nova Mon 2013 from the three
ANS telescopes is listed in Table~2 and presented in the Figure 2.

  \begin{table}
    \centering      
     \caption{Our merged photometry of Nova Mon 2012 (the table is published
  in its entirety in the electronic edition of this journal.  A portion is
  shown here for guidance regarding its form and content).}
    \includegraphics[width=8.4cm]{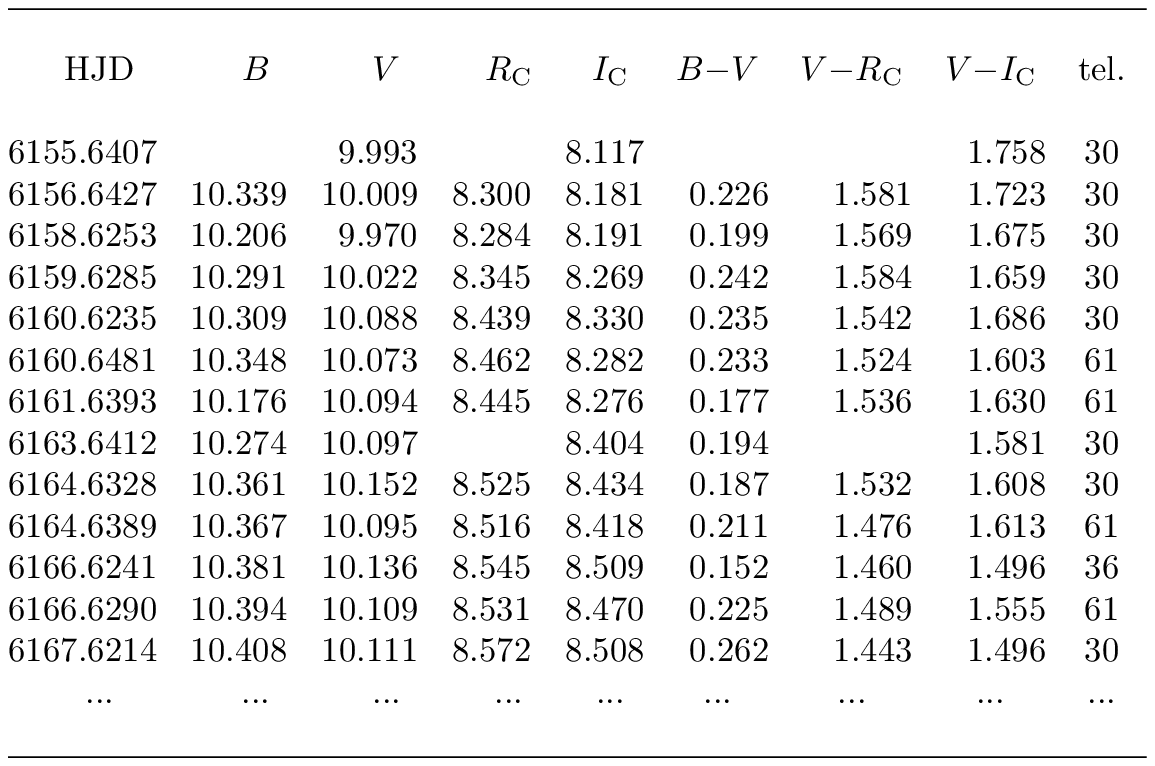}
     \label{tab2}
  \end{table}    

\section{Photometric evolution}

   \begin{figure*}
    \centering   
    \includegraphics[width=14cm]{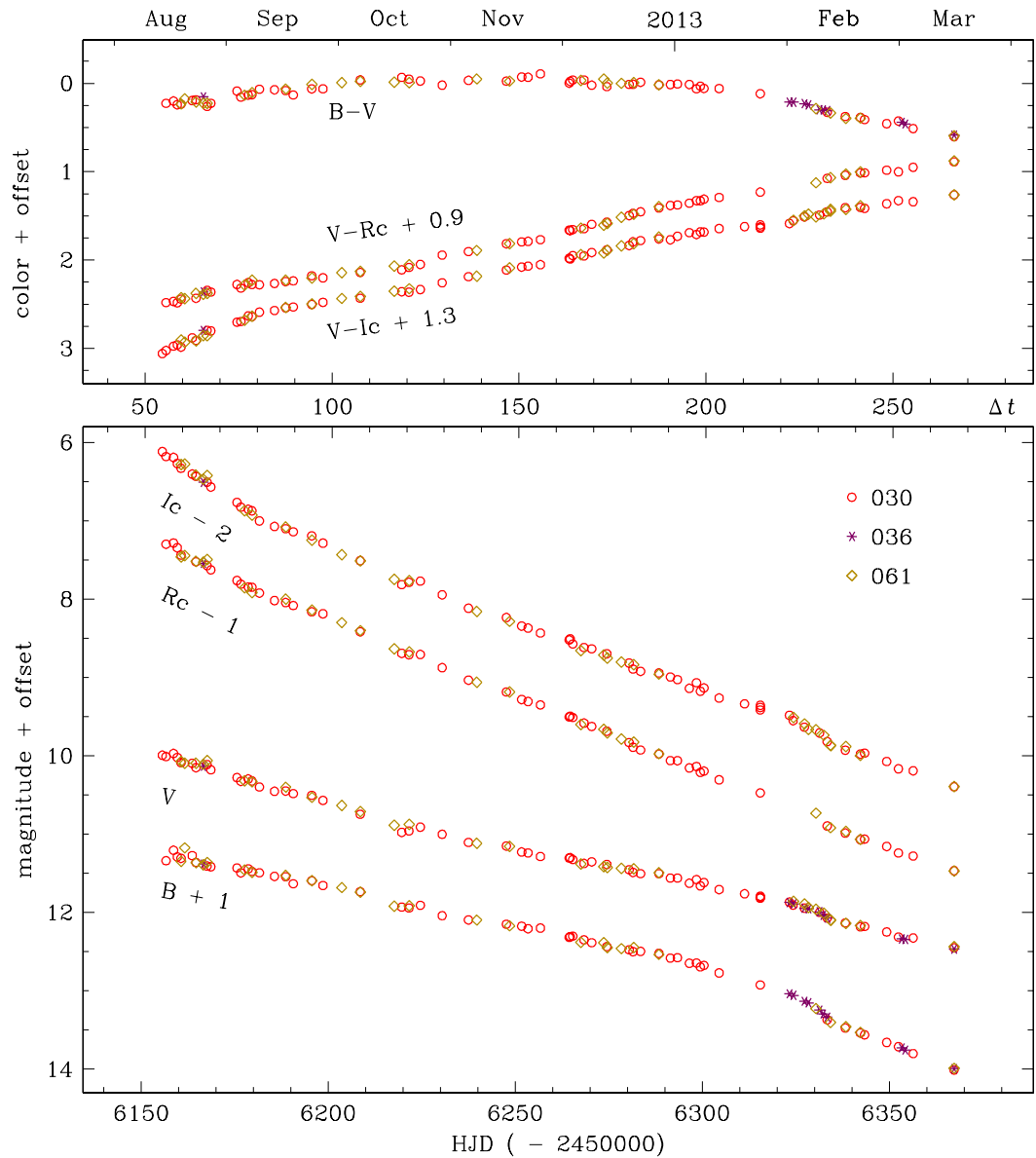}
     \caption{$B$$V$$R_{\rm C}$$I_{\rm C}$ photometric evolution of Nova Mon
     2012 as recorded with ANS Collaboration telescopes N. 30, 36 and 61. 
     The larger scatter of the first points is caused by the nova being at
     that time observable only very low on eastern horizon with dawn
     twilight already underway. The noise in the rest of the light-curve is
     primarily driven by the sampling along the 7.1 hr orbital modulation.
     The {\em knee} occouring around JD 2456330 is evident.}
     \label{fig2}
  \end{figure*}

\subsection{Broad bands}

The photometric evolution of Nova Mon 2012 shown in Figure~2 is one of the
smoothest ever recorded, at least until the development of the {\it knee} at
the beginning of February 2013.  The very low amplitude {\it noise}, which
is visible at an amplitude appreciably larger than the $\sim$0.007 mag total
error budget of the measurements, is due to the superimposed 7.1 hr period
modulation that will be discussed in the next section and which appear to be
orbital in origin.  The earliest points in the light-curve of Figure~2 are
noisier because the nova was at that time observed very low on the horizon,
in pre-dawn twilights soon after emergence from the conjunction with the
Sun.

At the time of the discovery at optical wavelengths, on day +48 from initial
appearance in gamma-rays, the nova was already well into its nebular stage,
with [OIII] of an intensity similar to H$\beta$ (Munari et al. 2012a).
Fitting with synthetic or observed light-curves does not constrain the
magnitude of the unobserved maximum.  As shown in Figure~2, the nebular
portion of the light-curve is close to a straight line and any fit would slip
along it. We can however work out some lower limit to brightness at maximum.
The transition from optically thick to optically thin conditions occurs
about $\Delta$mag=3.5 mag below maximum brightness. Thus the characteristic
time $t_3$ (the time required to decline by 3 mag below maximum brightness)
must have been significantly shorter than 48 days. Adopting $E_{B-V}$=0.30,
a distance of 1.5 kpc as constrained by the progenitor (see sect.6 below),
and averaging over Schmidt (1957), Pfau (1976) and de Vaucouleurs (1978)
relations for absolute magnitude at maximum as function of $t_3$ (cf. also
della Valle 1991), the apparent magnitude of Nova Mon 2012 at maximum
brightness would have been $V$= +2.8, +3.6, +4.0 or +4.3 mag for $t_3$=10,
20, 30 or 40 days, respectively, well within the nake-eye range.

The most striking feature of the broad-bands light-curves in Figure~2 is the
{\it knee} that started to develop around day +230.  It is simultaneously
visible in $B$,$V$ and $I_{\rm C}$, but not in $R_{\rm C}$.  The flux in the
latter band is largely dominated by the emission in H$\alpha$, which
accounted for 76\% of the whole flux in $R_{\rm C}$ band at that time.  The
absence of a visible knee in the $R_{\rm C}$ light-curves suggested that the
emission in H$\alpha$ continued to evolve smoothly, ignoring the occurrence
of the knee.  Such a sharp discontinuity in the light-curve of novae was
already observed at the time of switch off of the X-ray super-soft source
(attributed to the end of the nuclear burning at the surface of the WD) in
the recent outbursts of V407 Cyg (Munari et al.  2011a) and U Sco (Munari et
al.  2010), marking the sudden end of continued energy input by the central
source to the expanding ejecta.  At the same time of the appearance of the
knee, the SSS component in Nova Mon 2012 was rapidly declining (Page et al. 
2013a,b), even if this appeared to be a gradual though fast event and not a
switch off as observed in V407 Cyg and U Sco.  In addition, the
amplitude in magnitude of the knee in Nova Mon 2012 is much smaller than
observed in U Sco and V407 Cyg.  Still, the temporal coincidence is
striking, and suggests that also in Nova Mon 2012 the knee is related to a
major drop of energy input to the ejecta by the declining central source. 
The amplitude of the knee is about 0.2 mag in $B$ and a few hundreds in $V$
and $I_{\rm C}$.  To be detected, it requires very accurate observations,
tightly monitoring the advanced decline stages with the same telescopes
throughout the whole evolution, something that is rarely available for
novae.  We plan to contribute on future novae data of similar quality as
here presented, and with time we hope to build up a statistically
relevant sample to investigate the gross properties of such knees and their
correlation with global properties of novae, particularly at X-rays and
radio.  After the knee episode was over, Nova Mon 2012 settled on a decline
rate similar to that before the event.

  \begin{figure*}
    \centering   
    \includegraphics[width=13.5cm]{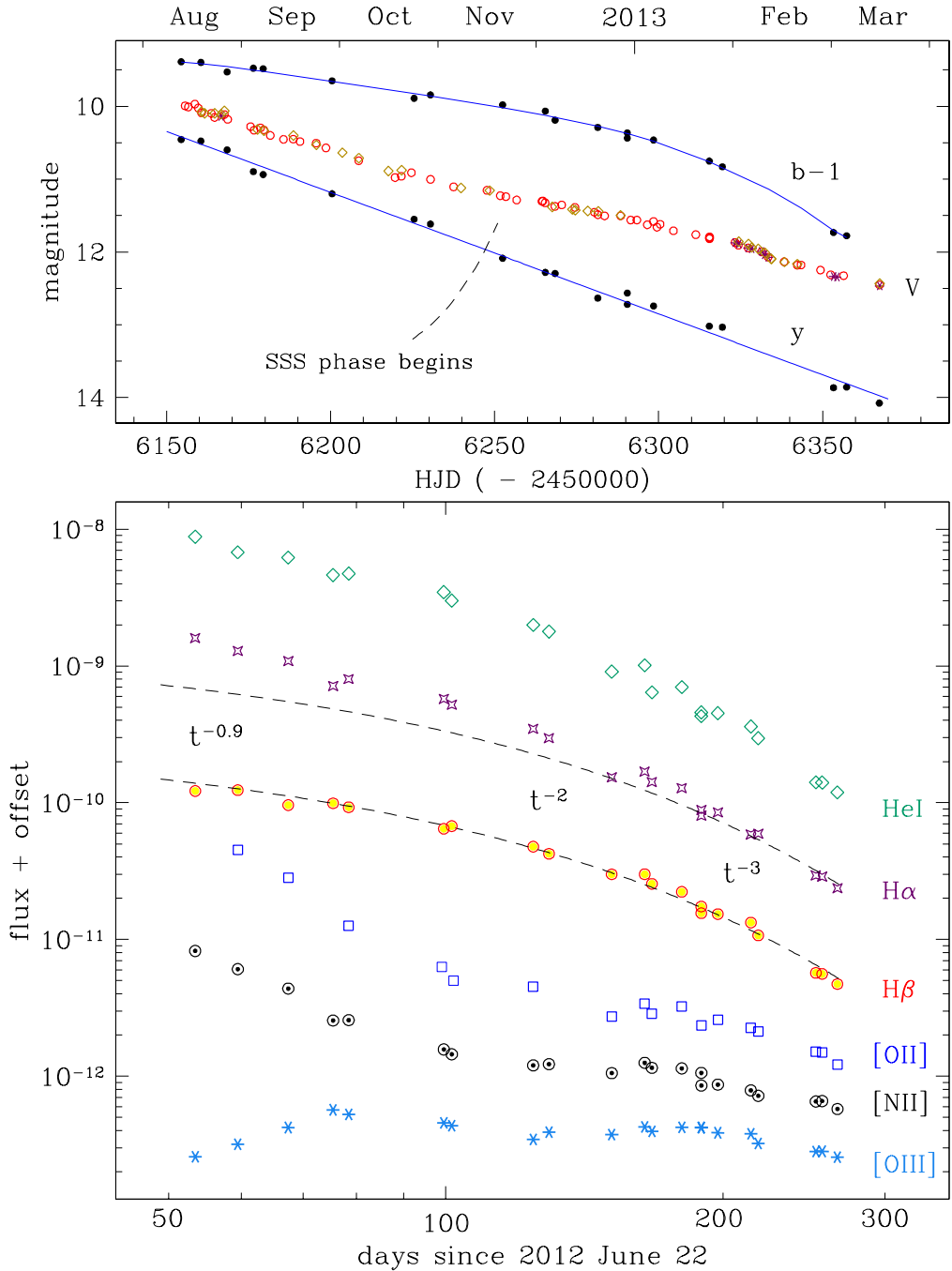}
     \caption{{\em Top panel:} photometric evolution of Nova Mon 2012 in
   Stromgren $b$ and $y$ filters compared to Landolt's $V$ band.
   {\em Bottom panel:} evolution of the integrated flux of some
   representative emission lines.  The dashed line is the flux (+ offset)
   recorded through the Stromgren $y$ band. Its $F \propto t^{-\theta}$ slope 
   is indicated at three different epochs.}
     \label{fig3}
  \end{figure*}

\subsection{Narrow bands}

Broad band optical photometry obtained with classical Johnson-Cousins
$U$$B$$V$$R_{\rm C}$$I_{\rm C}$ or SLOAN $u'$$g'$$r'$$i'$$z'$ systems
records simultaneously flux from the underlying continuum and the many
superimposed emission lines.  A detailed understanding of the physical
processes at work would require the flux of the continuum and the emission
lines to be measured separately.
 
A good filter to measure the intensity of the continuum with minimal
interference from emission lines is the $y$ band of the Stromgren system. 
Originally defined by a classical interference filter delivering a nearly
Gaussian passband, modern equivalent filters shape their transmission band
through application of multi-layered coatings, which offer a much higher
transmission and deliver near-rectangular passband profiles.  Figure~2 shows
the transmission profiles we have measured for the Stromgren $b$ and $y$
filters manufactured by Astrodon that we are just putting into
operation with ANS Collaboration telescopes to monitor future novae.
To anticipate and evaluate their results, we have integrated
their transmission profiles on our absolutely fluxed spectra of Nova Mon
2012 (cf Table~1) and by application of Bessell (2011) flux zero points
(Vega scale) transformed them into magnitudes.  The Stromgren $b$ and $y$
light-curves of Nova Mon 2012 are presented in the top panel of Figure~3
where they are compared to $V$ band replicated here from Figure~2.

As Figure~2 well illustrates, $y$ filter is placed in a wavelength region
deprived of significant emission lines during the nebular phase of a nova,
while [OIII] 4959, 5007 \AA\ and to a lesser extent H$\alpha$ emission lines
contribute heavily to the flux recorded through $V$ band.  This accounts for
the initially nearly constant offset between $y$ and $V$ bands, and their
bifurcation at later times. The bifurcation between $y$ and $V$ light-curves
begun in mid November 2012, in coincidence with the first detection of a
super-soft component in Swift X-ray observations of Nova Mon 2012 (Nelson et
al.  2012b).  If $y$ filter accurately trace the continuum emission,
Stromgren $b$ filter is centered on the 4640 \AA\ blend of NIII lines
(produced by Bowen fluorescence through cascade recombination requiring He
II Ly$\alpha$ seed photons) and HeII 4686 + [NeIV] 4716 \AA\ blend, and the
light-curve that it traces is more sensitive to the evolution of these lines
than to the underlying continuum.

\subsection{Continuum and emission lines}

The lower panel of Figure~3 shows on a log-log plot the time evolution of
integrated flux of some representative emission lines of Nova Mon 2012.  For
reference purposes, the flux evolution of the underlying continuum (as
traced by the linear fit to the $y$ light-curve in the upper panel of
Figure~3) is over-plotted to the evolution of H$\beta$ and H$\alpha$.  The
local values of the $F \propto t^{-\theta}$ relation for the continuum are
marked.

By definition, during the optically thin nebular phase the emission in the
continuum comes mainly from e$^{-}$ recombining with ions, essentially
hydrogen and helium.  In a simplified geometry for the ejecta -- a
spherically symmetric shell with a high covering factor, finite thickness
and ballistic $v(r)$ expansion velocity --, the evolution of the $\theta$
index can be understood as the shifting balance between: (1) recombination
of ions and free electrons, proportional to the electron density, therefore
evolving like $t^{-3}$, and (2) reionization by the radiation field of the
hot central star.  The latter become less effective with passing time
because of ($i$) decline in the luminosity of the central star (and
eventually switching off of the nuclear burning at the surface), and ($ii$)
dilution of the radiation field caused by the continuing expansion of the
ejecta.  At early stages of the nebular phase observed in Nova Mon 2012,
reionization was more effective in keeping up with recombination and the
decline in flux was gentle ($F \propto t^{-0.9}$).  At later times the
energy input from the central star was becoming negligible in the overall
energy budget of the ejecta, and the decline in flux settled onto the
dilution time scale ($F \propto t^{-3}$).  Actual 3D structures of the
ejecta can add significant complication to this simplified scenario. 
Bipolar lobes, equatorial wrists, polar caps, polar jets, prolate extended
volumes etc.  are structures that are normally encountered in high
resolution spatial imaging of nova ejecta or 3D modeling of their high
resolution emission line profiles (eg.  Woudt et al.  2009, Ribeiro et al. 
2009, Munari et al.  2011b, Chesneau et al.  2012).  Such spatial structures
have different densities and expansion velocities, are illuminated
differently by the central star, turn optically thin at different times, and
contribute to the overall brightness by different proportions.  It is worth
noticing that a highly bipolar structure of the ejecta of Nova Mon 2012 is
implied by high resolution radio imaging (O'Brien et al.  2012, Chomiuk et
al.  2013) and by modeling of the high resolution profiles of its emission
lines (Ribeiro et al.  2013).

The H$\beta$ emission line is produced by hydrogen recombination, which is
proportional to the electron density in the same way it is the emission in
the continuum.  It is therefore not surprising that the temporal evolution
of the continuum and H$\beta$ fluxes coincide in Figure~3.  On the other
hand, H$\alpha$ settled onto the continuum/H$\beta$ temporal evolution only
at later times, initially showing an extra flux component that gradually
disappeared.  This extra flux is probably related to ejecta's high optical
depth in hydrogen Ly$\beta$.  During the continuing absorption and
re-emission of Ly$\beta$ photons, a fraction of them is splinted into
H$\alpha$ and Ly$\alpha$ photons, and these extra H$\alpha$ photons are
in addition to those produced by hydrogen recombination.  The required high
optical depth in hydrogen Ly$\beta$ is confirmed by the strength of OI 8446
\AA, which is pumped by fuorecence from hydrogen Ly$\beta$.  The intensity
of OI 8446 \AA\ declined in parallel with the demise of the extra flux in
H$\alpha$.

  \begin{figure}
    \centering   
    \includegraphics[width=8.5cm]{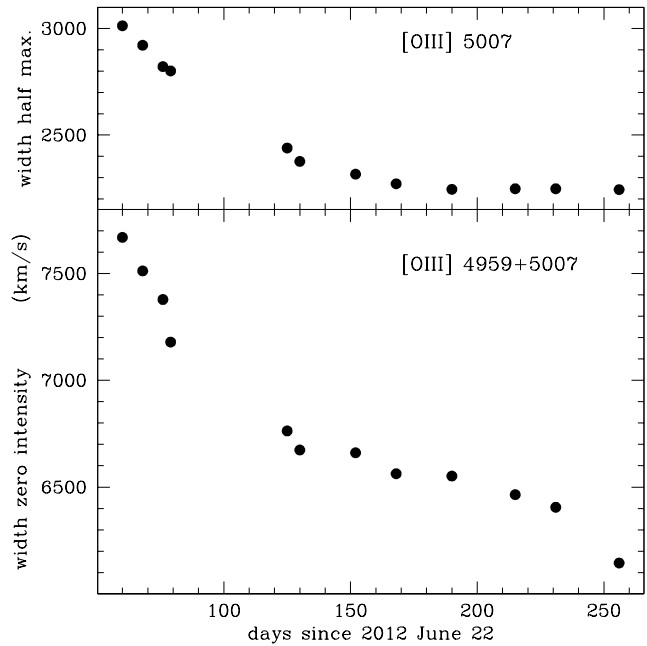}
     \caption{Time evolution of the width at half maximum of [OIII] 5007 \AA\
     emission line and of the width at zero intensity of [OIII] 4959+5007 
     blend.}
     \label{fig4}
  \end{figure}   

Interestingly, HeI 5876 shows in Figure~3 a flux evolution similar to that
seen in H$\alpha$, with a decline faster than that of H$\beta$ and the
nebular continuum.  The reason is the shifting ionization balance between
HeII and HeIII, as traced by the flux ratio HeII 5412 / HeI 5876 which kept
increasing with time (we cannot use HeII 4686 to trace HeIII because this
line is blended with stronger [NeIV] 4716).  HeII 5412 \AA\ is produced during
recombination of HeIII, and HeI 5876 during recombination of HeII, with the
faster decline seen in HeI 5876 tracing the extra depletion of HeII ions
caused by the increasing ionization spreading through the ejecta.

The evolution of nebular lines are exemplified in Figure~3 by [NII] 5755,
[OII] 7325 and [OIII] 4959+5007 \AA\ transitions, which are characterized by
different critical electron densities for collisional de-excitation of their
meta-stable upper levels, $\log N_{e}^{crit}$=7.5, 6.8 and 5.8 cm$^{-3}$,
respectively.  In addition to the recombination, ionization from the central
source and ionization balance, the flux of nebular lines is also governed by
the time evolution of the local density.  Thanks to their higher critical
densities and lower ionization, [NII] 5755 and [OII] 7325 appear early in
the evolution of a nova, while [OIII] has to wait for more diluted
conditions and higher ionization of the ejecta to become prominent.  At the
time of Nova Mon 2012 discovery and first observations in Figure~3, [NII]
and [OII] were already rapidly declining in intensity, while [OIII] was
increasing and at an appreciably slower rate.  This is due to the combined
effects of (1) increasing ionization of the ejecta, with NII turning into
NIII and OII to OIII, (2) nebular lines forming in the outer, lower density
regions of the ejecta, where dilution of the reionizing radiation from the
central star is greater and the electron density ($\propto r ^{-3}$) decline
faster, and (3) the fact that parts of the ejecta had an electron density already
lower than the critical value for [NII] and [OII] (thus able to produce the
5755 and 7325 \AA\ emission lines) but still higher than the critical value
for [OIII] (thus suppressing the formation of 4959+5007 \AA\ doublet).

A long lasting plateau is characteristic in Figure~3 of the late temporal
evolution of nebular lines, in particular for [OIII].  The plateau is the
result of two opposing effects that nearly counter-balance each other.  On
one side, the continuing expansion, by reducing the electron density,
diminishes the number of recombinations per unit time and therefore the
number of emitted photons.  On the other side, the continuing expansion
shifts inward in mass the boundary of critical density for collisional
de-excitation of their meta-stable upper levels, thus allowing to an
increasing fraction of the ejecta to emit the nebular lines.  This has a
kinematical consequence on the formation of the nebular line profiles. 
Diminishing the contribution from outer and faster moving layers of the
ejecta, at the same time as emission from the inner and slower moving layers
is progressively added, causes a reduction in the width of the nebular lines
and a suppression of their extended wings.  This is precisely what is
observed, as illustrated in Figure~4 where the width at half maximum of
[OIII] 5007 \AA\ and the width at zero intensity of the [OIII] 4959+5007
\AA\ doublet are plotted as function of time.

\section{Orbital modulation}

  \begin{figure*}
    \centering   
    \includegraphics[width=17.6cm]{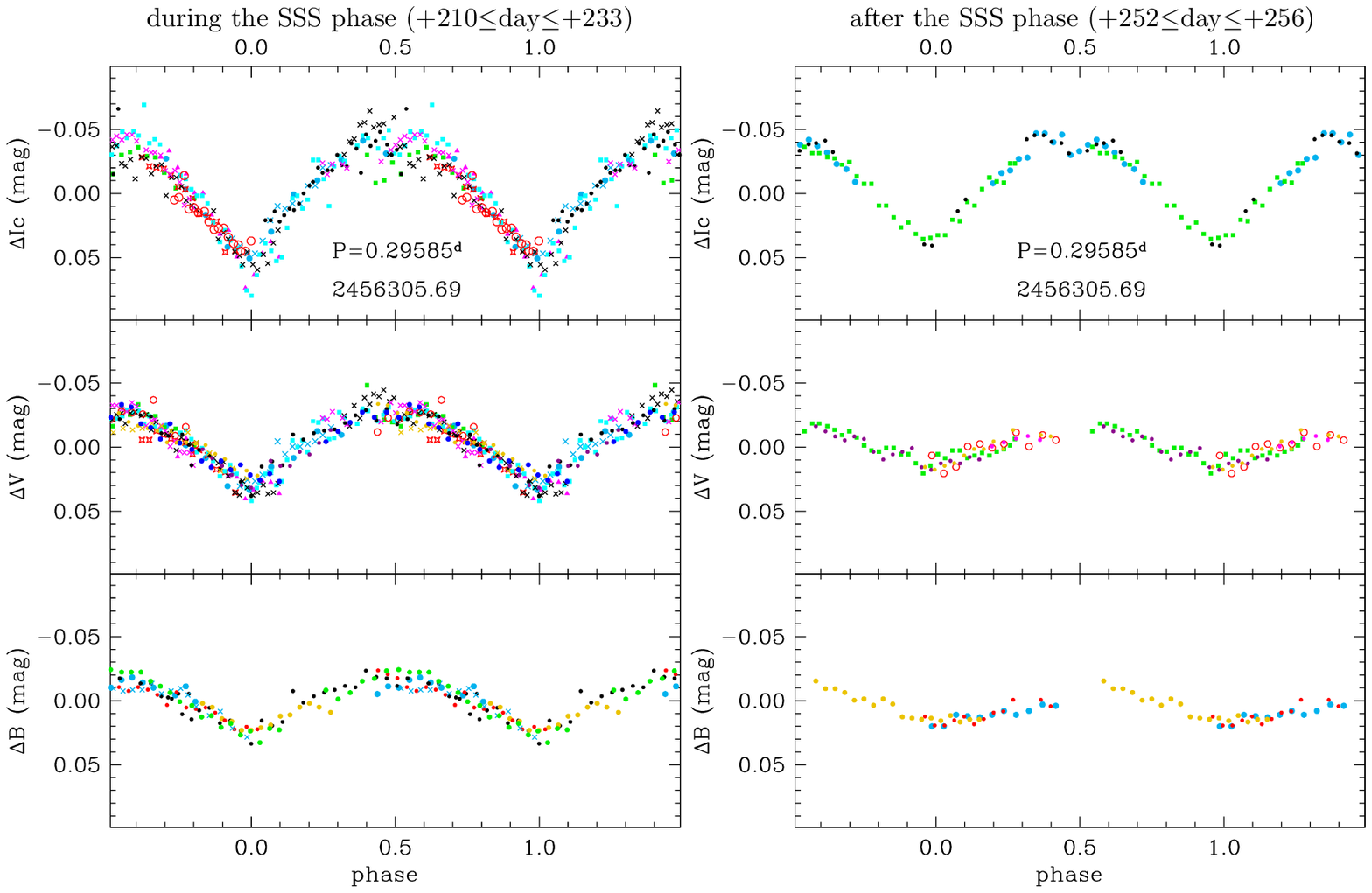}
     \caption{Effect of the 7.1 hr orbital period on the brightness of Nova
     Mon 2012.  {\em Left panel:} modulation observed while the super-soft
     source was still ``on'' (data from long monitoring runs on 2013 January 18,
     30 and 31, and February 1, 3, 4, 7, 8, 9, 10).  {\em Right panel:}
     modulation observed after the super-soft source switched ``off'' (data
     from 2013 March 1, 2, 3, 4, and 5 monitoring runs).}
     \label{fig5}
  \end{figure*}   

First suggestion of an orbitally modulated X-ray variability of Nova Mon
2012 was put forward by Orio et al.  (2012) based on their 7-hour long
integration to obtain a high-resolution X-ray grating spectrum with Chandra
during the SSS phase.  Osborne et al.  (2012) confirmed the large and rapid
variability in X-rays of Nova Mon 2012, and found in the ultraviolet Swift
data a marked 0.2957 ($\pm$0.0007) days periodicity with an amplitude of
$\sim$0.05 mag.  A similar periodicity (0.2957$\pm$0.0015 days) modulated
the X-ray emission.  Folding the X-ray and ultraviolet data shows a roughly
sine-like modulation, which had a common phasing in the two wavebands. 
Osborne et al.  (2012) attributed the modulation to a partial eclipse of
extended emission by an accretion disk rim which is raised at the point of
impact of the stream from the secondary star, implying that accretion disk
exists at the time of super-soft emission.  Wagner et al.  (2013) confirmed
the presence of a 0.2956 ($\pm$0.0010) day periodicity in $I_{\rm C}$
photometry obtained during 2013 January, sinusoidal in shape and with 0.13
mag full amplitude, and argued for it to be the orbital period of the
system.  A slightly longer period, 0.2965 $\pm$0.0072 days, was derived by
Hambsch et al.  (2013) from optical observations obtained in early February
2013, with full amplitudes of 0.06 mag in $B$, 0.08 in $V$, 0.10 in $R_{\rm
C}$ and 0.13 in $I_{\rm C}$.

With the same three ANS telescopes used for to construct the long term
photometric evolution of Figure~2 and Table~2, we carried out long 
nightly observing runs of Nova Mon 2012 to characterize such 7.1 hr periodicity.  
We splinted the campaign in three parts. 

\subsection{Before the SSS phase}

Nelson et al. (2012b) reported that Swift observations show Nova Mon 2012 to
enter the super-soft X-ray source (SSS) phase on Nov 18, 2012, i.e. on day
+150 from the initial Fermi-LAT detection. This marks the time when the
ejecta and the material/wind closer to the WD begin to get transparent to 
soft X-rays copiously emitted by the WD envelope where stable H-burning 
is undergoing (Krautter 2008).

The data listed in Table~2 and obtained as part of the normal nova
monitoring before the start of the SSS phase on day +150, were searched for
presence of the 7.1 hr after removing from them the nova decline.

Even if the data show a variability with an amplitude similar to that seen
at later times, no periodicity was found, in particular around 7.1 hr. It
seems that the same optically thick material close to the WD that was
blocking the soft X-rays prior to day +150 was also effective in preventing
optical detection of the 7.1 hr modulation, as if the central binary was
orbiting interior to the optical pseudo-photosphere. This is consistent with
the still modest ionization of the ejecta observed before day +150. High
ionization [OIII], [NeIV], [FeVII], etc. emission lines either appeared or
started to grow significantly in intensity only with the beginning of the
SSS phase.

\subsection{During the SSS phase}

During the SSS phase, long photometric runs lasting many consecutive hours
were carried out on 2013 January 18, 30, 31 and February 1, 3, 4, 7, 8, 9, 10,
i.e. covering the period from day +210 to +233.  After subtracting the mean
value of the magnitude for the given night, we have Fourier searched the
residuals and found a strong periodicity following the ephemeris:
\begin{equation}
Min = 2456305.69 + 0.29585(\pm0.00002)\times E
\end{equation}
The zero epoch is the same as defined by Osborne et al.  (2012) and Wagner
et al.  (2013), and our period has an accuracy higher by two orders of
magnitude. The phased data are presented in the left panel of
Figure~5. The near-sinusoidal variability has a total amplitude of
0.044($\pm$0.002) mag in $B$, 0.059($\pm$0.003) in $V$ and 0.094($\pm$0.004)
in $I_{\rm C}$ band. The exact value of the amplitude and its change with
time is a critical ingredient in understanding its origin, and therefore it
is important to accurately establish it. Our amplitude is smaller than
reported by Wagner et al.  (2013) and Hambsch et al.  (2013). The reason is
that they did not transform their photometry from the instantaneous local
photometric system to the standard one through rigorous application of the
instantaneous color equations calibrated for each frame on the photometric
sequence composed by many different stars well distributed in color and
magnitude. They instead limited to measure the magnitude of Nova Mon 2012
differentially with respect to one or two randomly chosen field stars.
During long monitoring runs, the height above the horizon of Nova Mon 2012
changed by large margins, and the difference in the energy distribution of
the nova with respect to the pair of field stars selected for the
differential photometry caused large and uncontrolled drifts in the zero
points. To test this scenario, we re-reduced our $I_{\rm C}$ band data and
measured Nova Mon 2012 differentially against a pair of blindly selected
field stars. As expected, this introduced a spurious additional
$\Delta$mag=0.085 variability, that combined with the real one leads to a
total $\Delta$mag=0.13, the same amplitude reported by both Wagner et al. 
(2013) and Hambsch et al.  (2013) for their $I_{\rm C}$ band observations.

\subsection{After the SSS phase}

Following the peak brightness reached on day +193, the SSS source started a
slow and monotonic decline, and by day +247 its brightness was reduced by 5
magnitudes, effectively disappearing around day +252 (2013 March 1; Page et
al. 2013b). To asses the origin and location of the sources responsible for
the 7.1 hr periodicity, in coincidence with the end of the SSS phase, we
started a new series of long observing runs on Nova Mon 2012, with the same
telescopes, observing procedures and photometric comparison sequence to
ensure the highest commonality with the previous observations. The new set
of observations extended from day +252 to +256, and the resulting
light-curves are shown on the right panel of Figure~5.

The new data adhere strictly to the same ephemeris (Eq.3) of the older ones.
The total amplitude of the variability reduced to 0.026($\pm$0.003) mag in
$B$ and 0.034($\pm$0.003) in $V$, whereas the amplitude of 0.090($\pm$0.003)
in $I_{\rm C}$ band remained essentially unchanged. A clear secondary
minimum is visible at phase 0.5. It could be present also during the SSS
phase (left panel of Figure~5), but available data are inconclusive.

The lightcurves of both panels of Figure~5 are particoularly smooth.
Any flickering possibly associated with the mass transfert, if accretion was
present at that time, contributed a negligible fraction to the total
flux compared to the expanding ejecta and the irradiated donor star.  The
larger dispersion of points around phase 0.5 in the $I_{\rm C}$ lightcurve
of the left panel of Figure~5 could be an artifact. In fact, by chance, it
happened that the last or the first observation of almost every observing run
was obtained in the $I_{\rm C}$ band and near orbital phase 0.5.

\subsection{Central binary}

The great stability in phasing of the 7.1 hr modulation irrespective of the
SSS phase proves beyond doubt its orbital origin. The presence of the
secondary minimum cleans the board from the possibility that the actual
orbital period is twice longer.

The donor, an early K-type dwarf star (cf. next section), fills its
Roche lobe, and the consequent ellipsoidal distortion would modulate the
lightcurve with two maxima and two minima per orbital cycle.  The orbiting
hot WD irradiates the facing side of the K star, that reprocesses at optical
wavelengths the excess radiation.  The resulting lightcurve is essentially
sinusoidal, with a maximum when the WD transits at inferior conjunction and
the irradiated side of the K star is in full view, and a minimum when the WD
is at superior conjunction and the irradiated side of the K star is hidden
from view.  The observed lightcurve is the combination of the two effects
superimposed, the irradiation being more important at shorter wavelengths
and the ellipsoidal distortion emerging at longer ones because there peaks
the energy distribution of the donor star.  The emergence at longer
wavelengths of a secondary minimum around phase 0.5, as seen for Nova Mon
2012 in Figure~5, is regularly observed in symbiotic binaries, where a cool
giant that fills its Roche lobe is irradiated by a very hot WD companion. 
The shape of $B$, $V$, and $I_{\rm C}$ lightcurves presented by Mikolajewska
et al.  (2003) for the high inclination symbiotic binaries YY Her and CI Cyg
are very similar to those presented in Figure~5 for Nova Mon 2012, which
also is a high inclination system seen close to edge-on conditions according
to Ribeiro et al.  (2013).

When the SSS phase ended and the WD stopped burning hydrogen at the surface
(right panel of Figure~5), its luminosity, temperature and radiation output
begun to decline.  This had the consequence that a lower amount of WD
radiation impinged on facing side of the K star companion, which
consequently declined in brightness, and this in turn meant a lower
visibility of the orbitally induced modulation against the overwhelming
glare of the circumstellar ejecta.  The large reduction in $B$ and $V$
amplitude observed between the two observing epochs in Figure~5 is thus due
to the reduced irradiation of the K star.

It may also be possible that an accretion disk, if present around the WD as
suggested by Osborne et al (2012), could be (partially) occulted at primary
minimum, somewhat contributing to the observed overall decline in
brightness.  Its transit over the irradiated side of the donor star around
phase 0.5 could equally contribute to the observed secondary minimum.

Observations similar to those presented in Figure~5, and carried out after
Nova Mon 2012 will re-emerge from conjunction with the Sun in September
2013, will be able to quantify the relative contribution and importance of
irradiation, ellipsoidal distortion and disk occultation in shaping the
orbitally modulated lightcurve. By then, in fact, the glare of the nova
ejecta and the irradiation of the donor star will be greatly reduced,
allowing to search for signature of disk occultation.

\section{The progenitor}

Banerjee et al. (2013) suggested that a common mechanism to produce
gamma-rays must be at work in all the novae detected by Fermi-LAT and in RS
Oph too, namely diffusive acceleration of particles across the shock front
formed as a consequence of the high velocity ejecta from the nova colliding
with a slow moving, extended and massive envelope, presumably created by
heavy mass loss from the companion.  Banerjee et al.  (2013) noticed that:
(1) infrared data for the progenitor suggest large pre-outburst variability:
on 2000 Feb 29, 2MASS measured the progenitor at J=15.759, H=15.231 and
K=15.036, whereas on October 2008 UKIDSS measured J=16.263, H=15.709 and
K=15.420, and (2) WISE observations in March 2010 detected the progenitor in
all four bands W1, W2, W3 and W4 (at 3.4, 4.6, 12 and 22 microns
respectively) with magnitudes of 14.732, 14.740, 10.831 and 8.720
respectively.  The large W2-W3 and W3-W4 colors are indicative of a large
mid-IR excess that is likely to be associated with a dusty extended
envelope. According to Banerjee et al. (2013), all this would imply that a
cool giant is present in Nova Mon 2012, and that it is loosing mass at a
high rate, leading to dust condensation in the circumstaller environment.

  \begin{figure}
    \centering   
    \includegraphics[width=8.5cm]{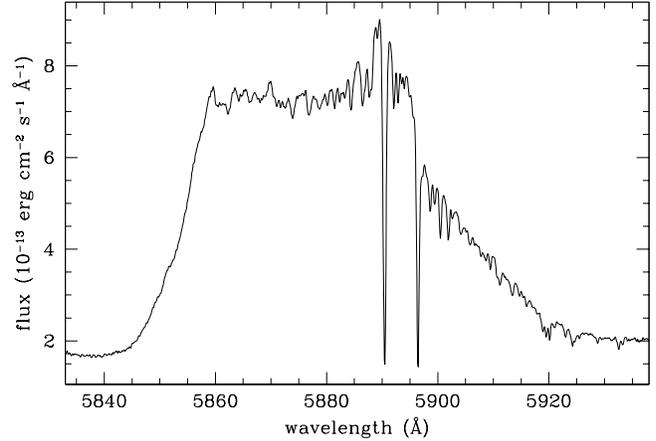}
     \caption{Interstellar NaI doublet 5890, 5896 \AA\ superimposed on
     the HeI 5876 emission line on the Echelle spectrum of Nova Mon 2012 for
     5 Sept 2012.  The many sharp lines on the red wing of the broad HeI
     line profile are due to telluric absorptions.}
     \label{fig6}
  \end{figure}   

  \begin{figure}
    \centering   
    \includegraphics[width=8.5cm]{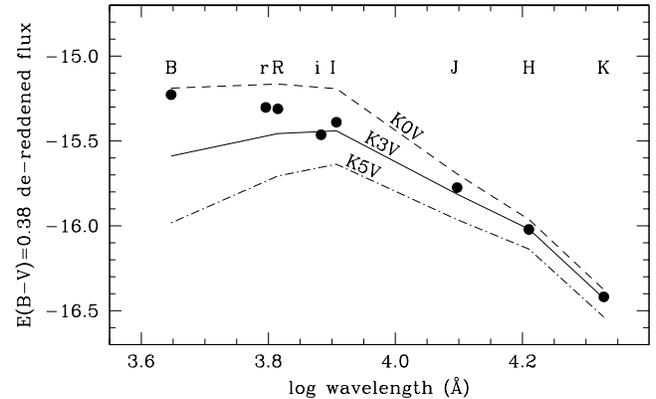}
     \caption{De-reddened ($E_{B-V}$=0.38) energy distribution of the
     progenitor (dots) compared with the energy distribution of early K-type
     main sequence stars at 1.5 kpc distance.}
     \label{fig7}
  \end{figure}   

There are observational evidences that argue against this model for the
progenitor of Nova Mon 2012.

The first one is 7.1 hr orbital period.  The Roche lobe associated with such
an orbital period is far too small to allocate room for a giant, a sub-giant
or even a turn-off star. There is always the possibility - not forbidden by
Nature - that the progenitor is a triple system, where an interacting P=7.1
hr close binary is orbited at a greater distance by a mass loosing cool
giant companion. However, such ad hoc hierarchical structure should be
exceeding rare to be statistically worth considering.

  \begin{table}
    \centering      
     \caption{Radius of the Roche lobe of the donor star in Nova Mon 2012
     for an orbital period 0.29588 days, and various WD and donor star masses.}
    \includegraphics[width=5.4cm]{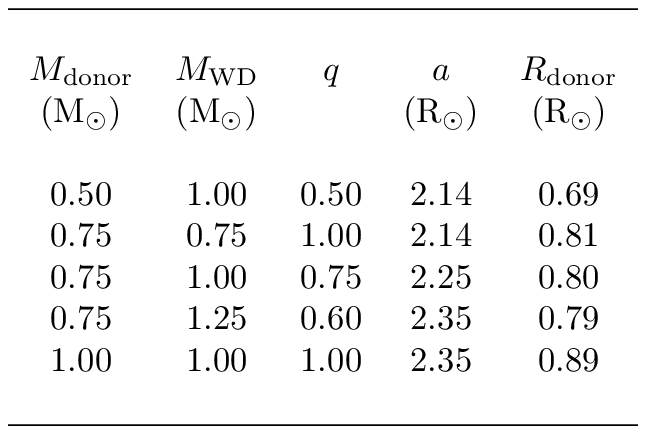}
     \label{tab3}
  \end{table}    

The second one is the faint magnitude in quiescence and the proximity of the
progenitor.  Nova Mon 2012 suffers from low reddening, as suggested by
the blue $B$$-$$V$ color that the nova displayed (cf.  Figure~2) and the
intensity of the interstellar lines.  From an Echelle spectrum obtained
immediately after the announcement of the nova, Munari et al.  (2012a)
preliminary estimated a reddening $E_{B-V}$=0.30 from the equivalent width
of very sharp, single component NaI 5890, 5896 \AA\ interstellar lines.  As
listed in Table~1, we have now at our disposal many more Echelle spectra to
accurately measure the interstellar lines.  An example is presented in
Figure~6.  The mean value (and error of the mean) for the equivalent with of
NaI 5890 \AA\ line is 0.536$\pm$0.006 \AA.  Adopting the Munari and Zwitter
(1997) calibration, this turns into a reddening $E_{B-V}$=0.38$\pm$0.01,
equivalent to an extinction of $A_V$$\sim$1.2 for a standard $R_V$=3.1
reddening law (Fitzgerald 1999).  According to Feast et al.  (1990) and the
literature that they reviewed, traveling a mean distance of 1.2 kpc on the
galactic plane in the Solar neighborhood accumulates an extinction of
$A_V$$\sim$1.2, thus Nova Mon 2012 should not be much farer out.  The
heliocentric radial velocity we measured for the NaI interstellar lines is
$-$25.2$\pm$0.7 km/s which indicate a kinematical distance of about 1 kpc
when compared with the kinematical maps of Brand \& Blitz (1993).  Along the
line of sight to Nova Mon 2012, the radial velocity of the interstellar
material present a large discontinuity between 2 and 3 kpc from the Sun,
where it crosses Perseus spiral arm.  No interstellar line is observed in
the high resolution spectra of Nova Mon 2012 at heliocentric velocities
between 10 and 30 km/sec corresponding to the crossing of the Perseus spiral
arm by the line of sight.  From the amount of reddening and the radial
velocity of the interstellar material, the distance to Nova Mon 2012 is
constrained to be 1$\leq$$d$$\leq$2 kpc.  It is interesting to note that
preliminary analysis by Chomiuk et al.  (2012) of early radio data on the
nova indicate a distance of about $\sim$1.4 kpc.  At such short distance and
low extinction, the presence in Nova Mon 2012 of any giant or sub-giant star
would have boosted the quiescence magnitude at much brighter values than
observed for the faint progenitor.

In Figure~7 we have plotted the spectral energy distribution (SED, corrected for
$E_{B-V}$=0.38) of the progenitor of Nova Mon 2012.  The $B$, $R_{\rm C}$,
$I_{\rm C}$ and $r'$, $i'$ are mean values from multi-epoch values from USNO-B1
and IPHAS catalogs, respectively. Similarly, the $J$, $H$ and $K$ data are
the average of the values reported by 2MASS and UKIDSS. The epochs of all
these measurements are all different, and in view of the variability
displayed by the progenitor in quiescence (about 0.3 mag in a week according
to IPHAS), combining them all into a unique SED seems the only sensible
thing to do. The energy distribution is well fitted by that of a early K-type
main sequence star ($\sim$K2V) at a distance of 1.5 kpc ($J$,$H$,$K$ data from
Koornneef 1983, $B$,$R_{\rm C}$,$I_{\rm C}$ from Bessell 1990, absolute
magnitudes from Sowell et al.  2007).  The blue excess at $B$ and $R_{\rm
C}$ suggests the presence of an accretion disk in quiescence, in agreement
with evidence from a moderate H$\alpha$ emission seen in IPHAS pre-outburst
data by Greimel et al.  (2012).  Trying to fit with the energy distribution
of a red clump giant (K2III), the most abundant type of giant in the
Galactic disk, would push the distance to beyond the outer limit of the
Galactic disk and the reddening would be risen to such high values to be
completely incompatible with observations.  Even worst would be the results
of attempting the fit with M giants and supergiants of the types hosted in
V407 Cyg, RS Oph and T CrB.

We are then left with the inescapable conclusion that the donor star in Nova
Mon 2012 is a main sequence, early type K star.  Therefore the presence of a
cool, heavy mass losing giant proposed by Banerjee et al.  (2013) to explain
the gamma-rays observed from Nova Mon 2012 does not fit observational
evidences.  Nevertheless, the infrared excess they noted in the WISE data is
intriguing and should deserve further study.  The level of variability they
noted in the infrared is similar to the difference in magnitude comparing
magnitude estimates of the progenitor on the Palomar I and II survey images. 
The presence of an accretion disk, the high orbital inclination and possible
associated eclipses or partial occultations, the ellipsoidal distortion of
the Roche lobe filling companion, and the irradiation of the secondary,
could easily combine along the 7.1 hr orbital motion to account for the
variability noted in quiescence.

An early K-type main sequence donor star has a mass of $\sim$0.75 M$_\odot$
and a radius of $\sim$0.8 R$_\odot$ (Straizys \& Kuriliene 1981; Drilling \&
Landolt 2000).  Adopting the Eggleton (1983) expression for the radius of
the Roche lobe:
\begin{equation}
\frac{R_{\rm donor}}{a} = \frac{0.49 q^{2/3}}{0.6 q^{2/3} ~~+~~ ln(1 + q^{1/3})}
\end{equation}
\noindent
where $q$ is the mass ratio M$_{\rm donor}$/M$_{\rm WD}$, we found the
results listed in Table~3.  The condition of Roche lobe filling is well
matched by the donor star, and this in turn mutually reinforce the
confidence on P=0.29585 days to be the actual orbital period.

\section{Acknowledgements}
We would like to thank Stefano Moretti and Alessandro Maitan for their
assistance in the measurement of the photometric filter transmission
profiles.

\bsp

\label{lastpage}


\begin{thebibliography}{99}

\bibitem[\protect\citeauthoryear{Banerjee, Ashok, \& Venkataraman}{2012}]{2012ATel.4542....1B} Banerjee D.~P.~K., Ashok N.~M., Venkataraman V., 2012, ATel, 4542, 1 

\bibitem[\protect\citeauthoryear{Bessell}{1990}]{1990PASP..102.1181B} Bessell M.~S., 1990, PASP, 102, 1181 

\bibitem[\protect\citeauthoryear{Bessell}{2011}]{2011PASP..123.1442B} Bessell M.~S., 2011, PASP, 123, 1442 

\bibitem[\protect\citeauthoryear{Brand \& Blitz}{1993}]{1993A&A...275...67B} Brand J., Blitz L., 1993, A\&A, 275, 67 

\bibitem[\protect\citeauthoryear{Chesneau et al.}{2012}]{2012A&A...545A..63C} Chesneau O., et al., 2012, A\&A, 545, A63 

\bibitem[\protect\citeauthoryear{Cheung et al.}{2012}]{2012ATel.4310....1C} Cheung C.~C., Shore S.~N., De Gennaro Aquino I., Charbonnel S., Edlin J., Hays E., Corbet R.~H.~D., Wood D.~L., 2012, ATel, 4310, 1 

\bibitem[\protect\citeauthoryear{Chomiuk et al.}{2012}]{2012ATel.4352....1C} Chomiuk L., et al., 2012, ATel, 4352, 1 

\bibitem[\protect\citeauthoryear{Chomiuk et al.}{2013}]{Chomiuk2013} Chomiuk L., et al., 2013, ApJ, submitted

\bibitem[\protect\citeauthoryear{della Valle}{1991}]{1991A&A...252L...9D} della Valle M., 1991, A\&A, 252, L9 

\bibitem[\protect\citeauthoryear{de Vaucouleurs}{1978}]{1978ApJ...223..351D} de Vaucouleurs G., 1978, ApJ, 223, 351 

\bibitem[\protect\citeauthoryear{Drilling \& Landolt}{2000}]{2000asqu.book..381D} Drilling J.~S., Landolt A.~U., 2000, Allen's Astrophysical Quantities, 4th ed.,  A.N. Cox ed., AIP Press  Springer, 381

\bibitem[\protect\citeauthoryear{Eggleton}{1983}]{1983ApJ...268..368E} Eggleton P.~P., 1983, ApJ, 268, 368 

\bibitem[\protect\citeauthoryear{Feast, Whitelock, \& Carter}{1990}]{1990MNRAS.247..227F} Feast M.~W., Whitelock P.~A., Carter B.~S., 1990, MNRAS, 247, 227 

\bibitem[\protect\citeauthoryear{Fitzpatrick}{1999}]{1999PASP..111...63F} Fitzpatrick E.~L., 1999, PASP, 111, 63 

\bibitem[\protect\citeauthoryear{Fuhrmann et  al.}{2012}]{2012ATel.4376....1F} Fuhrmann L., et al., 2012, ATel, 4376, 1 

\bibitem[\protect\citeauthoryear{Fujikawa, Yamaoka, \& Nakano}{2012}]{2012CBET.3202....1F} Fujikawa S., Yamaoka H., Nakano S., 2012, CBET, 3202, 1 

\bibitem[\protect\citeauthoryear{Greimel et al.}{2012}]{2012ATel.4365....1G} Greimel R., Drew J., Steeghs D., Barlow  M., 2012, ATel, 4365, 1 

\bibitem[\protect\citeauthoryear{Koornneef}{1983}]{1983A&A...128...84K} Koornneef J., 1983, A\&A, 128, 84

\bibitem[\protect\citeauthoryear{Krautter}{2008}]{Krautter} Krautter J., 2008, in Classical Novae 2nd ed., M.F. Bode and A. Evans eds., Cambridge Astrophysics Series 43, pag. 232

\bibitem[\protect\citeauthoryear{Li \& Kong}{2012}]{2012ATel.4614....1L} Li K.~L., Kong A.~K.~H., 2012, ATel, 4614, 1 

\bibitem[\protect\citeauthoryear{Munari}{2012}]{2012JAVSO..40..582M} Munari U., 2012, JAVSO, 40, 582 

\bibitem[\protect\citeauthoryear{Munari}{2013}]{2013ATel.4709....1M} Munari U., 2013, ATel, 4709, 1 

\bibitem[\protect\citeauthoryear{Munari, Dallaporta, \& Castellani}{2010}]{2010IBVS.5930....1M} Munari U., Dallaporta S., Castellani F., 2010, IBVS, 5930, 1 

\bibitem[\protect\citeauthoryear{Munari, Dallaporta, \& Valisa}{2012a}]{2012ATel.4320....1M} Munari U., Dallaporta S., Valisa P., 2012a, ATel, 4320, 1 

\bibitem[\protect\citeauthoryear{Munari et al.}{2011a}]{2011MNRAS.410L..52M} Munari U., et al., 2011, MNRAS, 410, L52 

\bibitem[\protect\citeauthoryear{Munari et al.}{2011b}]{2011MNRAS.410..525M} Munari U., Ribeiro V.~A.~R.~M., Bode M.~F., Saguner T., 2011, MNRAS, 410, 525 

\bibitem[\protect\citeauthoryear{Munari et al.}{2012b}]{2012BaltA..21...13M} Munari U., et al., 2012b, BaltA, 21, 13 

\bibitem[\protect\citeauthoryear{Munari et al.}{2013}]{2013NewA...20...30M} Munari U., Henden A., Belligoli R., Castellani F., Cherini G., Righetti G.~L., Vagnozzi A., 2013, NewA, 20, 30 

\bibitem[\protect\citeauthoryear{Munari \& Moretti}{2012}]{2012BaltA..21...22M} Munari U., Moretti S., 2012, BaltA, 21, 22 

\bibitem[\protect\citeauthoryear{Nelson et al.}{2012a}]{2012ATel.4321....1N} Nelson T., Mukai K., Chomiuk L., Sokoloski J., Weston J., Rupen M., Mioduszewski A., Roy N., 2012a, ATel, 4321, 1 

\bibitem[\protect\citeauthoryear{Nelson et al.}{2012b}]{2012ATel.4590....1N} Nelson T., Mukai K., Sokoloski J., Chomiuk L., Rupen M., Mioduszewski A., Page K., Osborne J., 2012b, ATel, 4590, 1 

\bibitem[\protect\citeauthoryear{Ness et al.}{2012}]{2012ATel.4569....1N}  Ness J.-U., Shore S.~N., Drake J.~J., Osborne J.~P., Page K.~L., Beardmore A., Schwarz G., Starrfield S., 2012, ATel, 4569, 1 

\bibitem[\protect\citeauthoryear{O'Brien et  al.}{2012}]{2012ATel.4408....1O} O'Brien T.~J., et al., 2012, ATel, 4408, 1 

\bibitem[\protect\citeauthoryear{Orio \& Pa}{2012}]{2012ATel.4633....1O} Orio M., Pa B.~T., 2012, ATel, 4633, 1 

\bibitem[\protect\citeauthoryear{Osborne, Beardmore, \& Page}{2013}]{2013ATel.4727....1O} Osborne J.~P., Beardmore A., Page K., 2013, ATel, 4727, 1 

\bibitem[\protect\citeauthoryear{Page et al.}{2013a}]{2012ATel.4845....1P} Page K.~L., et al., 2013, ATel, 4845, 1 

\bibitem[\protect\citeauthoryear{Page et al.}{2013b}]{Page_sub} Page K.~L., et al., 2013, ApJ 768, L26

\bibitem[\protect\citeauthoryear{Pfau}{1976}]{1976A&A....50..113P} Pfau W., 1976, A\&A, 50, 113 

\bibitem[\protect\citeauthoryear{Ribeiro et al.}{2009}]{2009ApJ...703.1955R} Ribeiro V.~A.~R.~M., et al., 2009, ApJ, 703, 1955 

\bibitem[\protect\citeauthoryear{Ribeiro et al.}{2013}]{2013ApJ...768...49R} Ribeiro V.~A.~R.~M., Munari U.,  Valisa P., 2013, ApJ, 768, 49 

\bibitem[\protect\citeauthoryear{Schmidt}{1957}]{1957ZA.....41..182S} Schmidt T., 1957, ZA, 41, 182 

\bibitem[\protect\citeauthoryear{Shahbaz et al.}{2003}]{2003ApJ...585..443S} Shahbaz T., Zurita C., Casares J., Dubus G., Charles P.~A., Wagner R.~M., Ryan E., 2003, ApJ, 585, 443 

\bibitem[\protect\citeauthoryear{Sowell et al.}{2007}]{2007AJ....134.1089S} Sowell J.~R., Trippe M., Caballero-Nieves S.~M., Houk N., 2007, AJ, 134, 1089 

\bibitem[\protect\citeauthoryear{Straizys \& Kuriliene}{1981}]{1981Ap&SS..80..353S} Straizys V., Kuriliene G., 1981, Ap\&SS, 80, 353 

\bibitem[\protect\citeauthoryear{Tatischeff  \& Hernanz}{2007}]{2007ApJ...663L.101T} Tatischeff V., Hernanz M., 2007, ApJ, 663, L101 

\bibitem[\protect\citeauthoryear{van den Bergh \& Younger}{1987}]{1987A&AS...70..125V} van den Bergh S., Younger P.~F., 1987, A\&AS, 70, 125

\bibitem[\protect\citeauthoryear{Varricatt et al.}{2012}]{2012ATel.4572....1V} Varricatt W.~P., Ehle J., Wold T., Banerjee D.~P.~K., Ashok N.~M., 2012, ATel, 4572, 1 

\bibitem[\protect\citeauthoryear{Waagen, Hambsch, \& Buil}{2012}]{2012CBET.3202....2W} Waagen E., Hambsch J., Buil C., 2012, CBET, 3202, 2 

\bibitem[\protect\citeauthoryear{Wagner, Woodward, \& Starrfield}{2013}]{2013ATel.4737....1W} Wagner R.~M., Woodward C.~E., Starrfield S., 2013, ATel, 4737, 1 

\bibitem[\protect\citeauthoryear{Warner}{1995}]{1995CAS....28.....W} Warner B., 1995, Cataclysmic Variable Stars, Cambridge Astrophys. Ser. 28

\bibitem[\protect\citeauthoryear{Woudt et al.}{2009}]{2009ApJ...706..738W} Woudt P.~A., et al., 2009, ApJ, 706, 738 

\end{thebibliography}
\end{document}